\documentstyle[sprocl,epsf]{article}

\newcommand{\eqa}{\begin{eqnarray}}
\newcommand{\ena}{\end{eqnarray}}
\newcommand{\eqar}{\begin{array}}
\newcommand{\enar}{\end{array}}
\newcommand{\eqn}{\begin{equation}}
\newcommand{\enq}{\end{equation}}

\font\blackboard=msbm10 scaled \magstep1
\font\blackboards=msbm7
\font\blackboardss=msbm5
\newfam\black
\textfont\black=\blackboard
\scriptfont\black=\blackboards
\scriptscriptfont\black=\blackboardss
\def\Bbb#1{{\fam\black\relax#1}}
\def\Bbb#1{{\bf #1}}
\def\yboxit#1#2{\vbox{\hrule height #1 \hbox{\vrule width #1
\vbox{#2}\vrule width #1 }\hrule height #1 }}
\def\fillbox#1{\hbox to #1{\vbox to #1{\vfil}\hfil}}
\def\ybox{{\lower 1.3pt \yboxit{0.4pt}{\fillbox{8pt}}\hskip-0.2pt}}
\def\comments#1{}

\def \II {II}
\def \IIa {IIa}

\def \Tr {{\rm Tr}}

\def\BC{\Bbb{C}}
\def\BP{\Bbb{P}}
\def\BR{\Bbb{R}}
\def\BZ{\Bbb{Z}}
\def\p{\partial}

\def\bra#1{{\langle}#1|}
\def\ket#1{|#1\rangle}
\def\vev#1{\langle{#1}\rangle}
\def\Dslash{\rlap{\hskip0.2em/}D}

\def\CA{{\cal A}}

\def\CN{{\cal N}}

\begin{document}
%\rightjustify{RU-99-07}

\title{TWO LECTURES ON D-GEOMETRY AND NONCOMMUTATIVE GEOMETRY}

\author{MICHAEL R. DOUGLAS\footnote{
Supported in part by DOE grant DE-FG05-90ER40559.}}

\address{Department of Physics and Astronomy, Rutgers University,\\
Piscataway, NJ 08855, USA\\
and\\
I.H.E.S., Le Bois-Marie, Bures-sur-Yvette France 91440}

\maketitle\abstracts{
This is a write-up of lectures 
given at the 1998 Spring School at the Abdus Salam ICTP.
We give a conceptual introduction to
D-geometry, the study of geometry as seen by D-branes in string theory,
and to noncommutative geometry as it has appeared in D-brane and
Matrix theory physics.
}

\section{Introduction}

Over the last few years, a new paradigm for thinking about fundamental
theories of physics has emerged, using the ideas of supersymmetry, 
effective theory, and duality.  The largest, perhaps all-encompassing
example, is called M theory, and subsumes the known superstring theories
and eleven dimensional supergravity.  We now have a fairly good understanding
of the moduli space of M theory compactifications with extended supersymmetry.

In contrast to some of the expectations raised by string theory
in the 80's, geometry as already
studied by the mathematicians: algebraic geometry, differential geometry,
and so on, has played a central role.  Partly this has been because of the
choice of questions.  One specifies an effective field theory by specifying
a field content and Lagrangian; on very general grounds the fields should
be thought of as a set of coordinates on some configuration space; the potential as a function; the kinetic term as a metric on this space.
This is true if we derive the theory in a conventional
Kaluza-Klein picture, and the geometry of this effective field theory has
much in common with that of the internal manifold.
It is just as true if we derive the theory from some generalization of
the Kaluza-Klein picture in which the internal space is not described
by conventional geometry, say because it is too small, because it is singular,
because quantum fluctuations are large, or perhaps
because our starting point is not geometrical.
Thus the language of differential geometry will always be useful; 
complex or K\"ahler geometry will be useful given the appropriate
supersymmetry, and so on.

Once we start studying the structure of the internal space itself we
might ask for an appropriate generalization of conventional geometry.
Although various proposals have been made, they
tend to focus on quantities which
appear in the effective theory and which are not local functions of the
geometry of the target space, such as the cohomology ring, or
volumes of supersymmetric cycles.  There is a good reason for this as
M theory is not local in the same sense as quantum field theory (though
the question of whether some definition of locality can be made is still
very much open; see \cite{dcurve} for a discussion of this).
Can we find analogs of the local invariants of conventional
geometry such as curvature, even in situations where conventional locality
does not apply ?

In these two lectures I will give two examples of the types of geometry
we might want to think about.  The first, geometry as seen by D-branes
or D-geometry, is directly motivated by the physics
of string theory, but is not yet well understood by either physicists or
mathematicians.

The problem of finding all backgrounds of string theory (say type \II\ for
definiteness)
is a generalization of the corresponding problem for supergravity,
which can be phrased as the problem of solving certain non-linear PDE's,
the equations of motion.  The generalization involves two parameters:
the string coupling $g_s$ (related to $\hbar$ as $\hbar=g_s^2$) and the
string length $l_s$.  Although no general and precise 
definition of the problem for $g_s\ne 0$ has been made (clearly this brings
in all of M theory) the case of $l_s\ne 0$ can be phrased
as the problem of classifying conformal field theories with an appropriate
amount of world-sheet supersymmetry.

A simple and general source of these is non-linear sigma models.
This starts from a target space metric but allows computing stringy
corrections to observables as an expansion in $(l_s/l_R)^2$ where $l_R$
is the curvature radius.  However the results cannot be directly thought
of as defining a new target space metric.

More recently D-brane physics allows us to introduce local probes into
our string backgrounds.  In particular a D$0$-brane moving in a sigma
model background sees a specific metric -- the ``D$0$-metric.''  This
is a uniquely defined local invariant of the string background, and
the first lecture will discuss it in some examples.

The second lecture will give a brief introduction
to noncommutative geometry, a mathematical
framework with many applications.  M theory is still one of its more
speculative applications but it appears that the noncommutative torus
does arise naturally in Matrix theory and D-brane physics, as we will
discuss, and provides a new geometrical picture of T-duality.

\section{What is D-geometry}

D-geometry is the study of the geometry of M theory compactification as
seen by a D-brane. 
The prototypical problem of D-geometry -- and the one we will discuss for
most of this lecture -- is: 
What metric does a D$0$-brane moving in the string background $M$ see?

By $M$ we mean
a conformal field theory which can be used in \IIa\ superstring theory;
by ``see a metric'' we want to say that the D$0$-brane has an effective
action of the form
\eqn\label{d-lagrangian}
S = {1\over g_s l_s} \int dt\ e^{-\phi} g^D_{ij}(X) \p_t X^i \p_t X^j 
+ \ldots 
\enq
and we want to determine the space parameterized by $X^i$ and $g_{ij}$ on
this space.

Already this point requires a bit of discussion as the D-brane is in itself
a quantum object and we should ask: how do we know this quantum mechanics
comes from quantizing a Lagrangian?
This should be true for sufficiently small $g_s$, but how small ?

Intuitively, we need to be able to form wave-packets which are small compared
to the curvature length $l_R$ of the metric, to be able to claim that we can
unambiguously recover the metric from the quantum mechanics.  This requires
$\hbar/p << l_R$ (which is the same as the condition for a WKB treatment
of (\ref{d-lagrangian}) to be justified).  The mass of the D$0$ is
$m=1/g_s l_s$ 
and (in our conventions with $\hbar=1$) we require $2 m E l_R^2 >> 1$ or
$E >> g_s l_s / l_R^2$.

This gives a minimum energy for the wave packets we need, but we also expect
(\ref{d-lagrangian}), as an effective action, to be valid only
up to some maximum energy.  An obvious criterion for integrating out
massive string
states is $E << 1/l_s$.  In many examples (as we will see)
there are also states associated with the curvature radius, and we also
require $E << l_R/l_s^2$.  This second condition will be stronger if
$l_R << l_s$, i.e. in the substringy regime where we might look for new
phenomena.

The conclusion is that we can think of the D$0$-brane as seeing a well-defined
metric if $l_R^2 >> g_s l_s^2$ and $l_R^3 >> g_s l_s^3 = l_{p11}^3$ the 
eleven-dimensional Planck length.  
As discussed in \cite{dkps}, one can think of the scale
$l_{p11}$ as defining a ``size'' of the D-brane, a scale where
its interactions become strong, and the moduli space picture breaks down.

By taking $g_s$ sufficiently small we
can explore the substringy regime, $l_R << l_s$.  In the following we will
do this by considering amplitudes from the sphere and disk world-sheets.

\subsection{General considerations from conformal field theory}

Given a conformal field theory defining a superstring compactification,
our first problem is to identify the
moduli space of boundary states corresponding to D$0$-branes.  The
conditions for a boundary state to correspond to a BPS state in space-time
are given in \cite{ooguri}, while
to decide that it is a D$0$-brane we should compute its RR charge.
D$p$-branes of different $p$ are generally related by dualities and
it is interesting to consider the same questions for $p\ne 0$, making
contact with mirror symmetry (e.g. see \cite{syz}), but that deserves
another lecture.

Massless fluctuations then correspond to boundary operators of dimension $1$;
call a basis for these $O_i$.
In the case that the moduli space is a manifold with local coordinates $x^i$,
we can choose the basis such that an insertion of $\int O_i$ corresponds
to the variation $\p/\p x^i$.  Although this will be true in the
examples we consider, it is not always true, in CFT terms because 
the operators $O_i$ can gain anomalous dimensions.
A counterexample is the minima of a potential such as $V =\phi_1^2
\phi_2^2$ such as would be found in the world-volume Lagrangian of a 
system of $0$-branes and $4$-branes.

The D$0$-brane metric will then be the Zamolodchikov metric on the 
moduli space.
We can define this as the two-point function on the disk,
\eqn\label{zammet}
g^D_{ij}(x) = \vev{ O_i(0) O_j(1) },
\enq
considered as a function on the moduli space.
The choice of two distinct points (here $0$ and $1$) on the boundary
is conventional.
Heuristically the relation between this and the effective action is clear as 
$g^D_{ij}$ determines the normalization of a state created by $O_i$.
To justify this from the string theory S-matrix one must consider
a four-particle scattering and relate this to the curvature $R[g^D]$, 
but this relation is standard in effective field theory.

Many supersymmetric string compactifications have been defined as non-geometric
conformal field theories (free fermion models, Gepner models, asymmetric orbifolds,
etc.), but some admit alternate geometric definitions (see \cite{greene} for an
overview).  It is an interesting question when this is possible, or to what extent
one can start with geometric models and by varying parameters (moduli or
even massive fields) reach all models.  Knowing the D$0$-metric would be a 
real help in finding geometric counterparts of these models.

\subsection{Non-linear sigma models}

A simple $M$ to consider is the supersymmetric non-linear sigma model
with background metric $g_{ij}(x)$.
This will be a conformal field theory if the beta function for the
metric vanishes.  This can be computed using the covariant background
field method.\cite{sigma}  One writes the fields as a sum $X=x_0+\xi$
where $x_0$ is a chosen background point and $\xi$ is a quantum fluctuation.
In Riemann normal coordinates the action
has an expansion whose coefficients are
tensors constructed from $g(x_0)$:
\eqn
S = {1\over l_s^2}\int d^2\sigma\ g_{ij}^{(0)}(x_0)\p\xi^i\p\xi^j 
+ R_{ijkl}[g^{(0)}] 
\p\xi^i\p\xi^k \xi^j \xi^l + \ldots
\enq
The leading term in the beta function then comes from a one-loop diagram
in which the $R$ interaction vertex has a single self-contraction,
producing a divergence proportional to the Ricci tensor.  Thus one
can obtain a conformal theory to this order by taking $g=g^{(0)}$ a
Ricci flat metric.

Working to higher orders, one obtains\cite{grisaru}
\eqn\label{beta}
0 = \beta_{ij} = R_{ij}[g^{\sigma}] + l_s^6 R^4_{ij}[g^{\sigma}] + O(l_s^8)
\enq
with $R^4_{ij}$ given in many references such as \cite{banksgreen}.
The condition is no longer Ricci flatness but it has been
shown that by adding {\it finite} corrections to $g^{(0)}$ at each
order in $l_s^2$ one can obtain a solution $g^{\sigma}$ of 
(\ref{beta}).\cite{nemsen}

Just as in flat space, the allowed boundary conditions on bosons 
are those with  $g_{ij}(X) \delta X^i \p_\sigma X^j=0$,
either Dirichlet or Neumann.  For Dirichlet boundary conditions
$X^i|_{\sigma=0}=x^i$, the marginal operators are 
$O_i = g^{(0)}_{ij} \p_\sigma X^j$
and the Zamolodchikov metric at leading order is just $(g^{(0)})_{ij}$.

We can proceed to compute $l_s$ corrections using the background field
method.  These have two origins a priori.  One type of correction comes
from bulk renormalization (closed string effects).  
Using the scheme of adding finite corrections to $g^{(0)}$,
these corrections produce the metric $g^{\sigma}_{ij}$, 
which is thus a better approximation to $g^D$.

In principle there can be additional finite corrections to the correlation
function (\ref{zammet}), which would naturally be tensors 
constructed from $g^\sigma(x_0)$. 
These are potentially important not just to get
the correct result but also for the following reason of principle.
The sigma model metric $g^{\sigma}_{ij}$ is not directly observable and
is in fact ambiguous.  In renormalizing the theory we need to make
choices for the finite parts of counterterms at each order in $l_s^2$;
a renormalization prescription.  A priori these could be
any local functionals of $g^{(0)}$.  This corresponds
to the possibility to make local field redefinitions in the resulting
target space supergravity theory.

On the other hand, the D-brane metric $g^D_{ij}$ is a metric on a
moduli space and as such is observable.  Its curvature will govern
the scattering of small fluctuations on the world-volume.  Although
there is an ambiguity of field redefinition in its computation as
well, this just corresponds to the freedom to choose a coordinate
system on the moduli space.  Of course if we considered other terms
in the world-volume Lagrangian, for example the gauge kinetic term,
we would have additional field redefinitions to deal with, but these
do not affect $g^D$.

Thus any additional finite terms contributing to $g^D$ will be scheme
dependent, in a way which exactly compensates the scheme dependence of
$g^\sigma$.  This suggests the possibility that there exists a preferred
renormalization scheme in which $g^D=g^\sigma$.  The best candidate for
this scheme is the usual minimal subtraction scheme (which subtracts
only poles in dimensional regularization) and so far it has been checked that 
in this scheme
finite terms do not appear at order $l_s^2$ and $l_s^4$.\cite{toappear}

The idea that $g^D=g^\sigma$ in a preferred scheme would
have a number of interesting consequences.  For one thing, it would
be determined by an ``equation of motion'' (\ref{beta}) expressed
as a series in local tensors.  Using $\CN=2$ perturbation theory,
another consequence is that the complex structure of $g^D$ would be
equal to that of the Ricci flat $g^{(0)}$.

In general observables can get non-perturbative corrections from 
world-sheet instantons as well; the metric on D$3$-brane
moduli space studied in \cite{syz} is an example.  One can show that at
least some D$0$-metrics do not;\cite{toappear} in a case where it does
the conjecture might be that
$g^D=g^\sigma$ to all orders in $l_s$ with finite computable corrections.
It would be an interesting question whether the result was governed
by an equation analogous to (\ref{beta}).

The nonlinear sigma model approach provides a universal definition of
$g^D$ (at least for all models connected to the large radius limit).
On the other hand we expect any new qualitative features to appear
in the substringy regime where the series expansion in $l_s^2/l_R^2$
is at present an impractical approach.
Rather, our knowledge of $g^D$
in the substringy regime comes entirely from examples.
Let us discuss a few.

\subsection{Orbifolds}

D-branes on orbifolds give a large set of substringy examples.
This has been a much studied subject (see \cite{dm} and papers which
cite it for a list of references)
and deserves a review of its own.  Here we focus on geometric aspects.

In general one defines string theory on an orbifold $M/\Gamma$ by
first adding to the theory twisted string sectors and then projecting
on $\Gamma$ invariant states.  For closed strings the twisted sectors
are closed only up to an element of $\Gamma$, while for open strings
the twisted sectors start and end on D-branes which will be identified
after we make the projection.  This problem is somewhat simpler than
that for closed strings as we can derive the complete D-brane
world-volume theory as a subsector of the theory of D-branes in $M$.

For $M=\BR^n$ this theory will be the usual maximally supersymmetric
super Yang-Mills so the entire construction is summarized in the
projection.  This can be written
\eqn\label{orbifold}
\gamma^{-1}(g) \phi \gamma(g) = R(g) \phi \qquad\qquad 
{\rm for\ all}\ g \in \Gamma,
\enq
where $\phi$ is the complete list of fields in the theory, $R(g)$ is
the action on space-time, and $\gamma(g)$ is an action of $\Gamma$
in some representation $r$ which we can freely choose -- string theory
consistency conditions allow any $r$.  The action of $R(g)$ can generally
be deduced geometrically -- for example if $g$ preserves the
world-volume and takes a world-volume point $z\rightarrow z'$, it will
act on a scalar field as $(R(g)\phi)(z)=\phi(gz)$, on a vector as
$(R(g)\phi)^i = R(g)^i_j \phi^j$ with $R(g)^i_j$ in the vector
representation, and so on.
Because the action is geometric it will preserve the supersymmetries
which are singlets under $\Gamma$, so we have the usual relation
between $\Gamma$ considered as a reduced holonomy group and unbroken
supersymmetry.  For example, $\Gamma\subset G_2$ would lead to three
dimensional gauge theories (and their dimensional reductions) with
$2$ supercharges.

In our D$0$-brane examples (\ref{orbifold}) leads to quiver gauge
theories.  Let $N=\dim r$; then
the gauge group is the commutant of $r$ in $U(N)$; if $r=\sum n_a r_a$
is the decomposition into irreducible representations $r_a$ this will
be
\eqn
G = \prod_a U(n_a) .
\enq
The matter will be a sum of bifundamentals, one for each singlet in
the product of representations
\eqn
R \otimes r \otimes r^* .
\enq
Another way to express this is in terms of the action
of the representation $R(g)$ in the representation ring:
if
\eqn\label{repring}
R \otimes r_a = \oplus_b N_{ab} r_b
\enq
there will be $N_{ab}$ bifundamentals in the representation 
$(n_a,\bar n_b)$.

For the D-brane scalars $X^i$, $R$ is the vector or fundamental
representation.  A natural choice for $r$, corresponding to a single
D-brane free to move on the orbifold, is the regular representation
$r_R = \sum_a (\dim r_a) r_a$
with basis vectors $\ket{g}$ for each $g\in\Gamma$ and matrix
elements
\eqa
\gamma(g)\ket{h} &\equiv \ket{gh} \cr
\bra{h} \gamma(g) &= \bra{g^{-1} h} \cr
\vev{g|h} &= \delta_{g,h} .
\ena
This choice expresses the intuition that the D-brane will have an ``image''
for each element of $\Gamma$.  It is easy to show that the moduli
space for this gauge theory includes $\BR^n/\Gamma$, by writing the
solution (for any constant $x^j$)
\eqn
X^i = \sum_{h\in\Gamma} \ket{h}\bra{h} R^i_j(h) x^j .
\enq
Since this is diagonal it will be a zero energy vacuum (this
condition for the $\CN=4$ theory includes the conditions for the
projected theories), while $\Gamma$ is realized by gauge
transformations.

There is a famous relation between discrete subgroups $\Gamma\subset
SU(2)$ and the ADE extended Dynkin diagrams.  It appears here as the McKay
correspondance -- for any $\Gamma$, the $r_a$ can each be associated
with nodes of a Dynkin diagram, and the links with non-zero $N_{ab}$
in (\ref{repring}).  Actually the best diagrammatic notation
(called quiver diagrams) is slightly different -- we write 
$N_{ab}/n_an_b$
{\it oriented} links between each pair of nodes $a$ and $b$.
This notation directly encodes the spectrum of
the resulting gauge theory, with each node now associated with the
gauge group $U(n_a)$ and each link with a complex matter field in
the bifundamental $(n_a,\bar n_b)$.

The $\BC^2/\Gamma$ theories will have eight supercharges and can be
written as dimensional reductions of $d=4$, $\CN=2$ theories.
The pseudoreality of the vector representation $R$
guarantees that each Dynkin diagram link will come once with
each of the two orientations, making up a hypermultiplet of $\CN=2$.
The $\Gamma=\BZ_2$ theories are then $U(n_1)\times U(n_2)$ gauge theory
with two hypermultiplets in the $(n_1,\bar n_2)$.

The $\BC^3/\Gamma$ theories can be written as dimensional reductions
of $d=4$, $\CN=1$ theories.  If $R$ is a complex representation of
$SU(3)$, one can get chiral gauge theories.  The simplest example is
$\Gamma=\BZ_3$ acting as $Z^i \rightarrow \exp 2\pi i/3\ Z^i$.  This
produces $U(n_1)\times U(n_2)\times U(n_3)$ gauge theory with
the chiral multiplet spectrum 
$3(n_1,\bar n_2) + 3(n_2,\bar n_3) + 3(n_3,\bar n_1)$.

\subsection{Resolved orbifolds}

What makes these constructions interesting realizations of substringy
geometry is the possibility of explicitly blowing up the orbifold
singularities in terms of the world-volume gauge theory.
It can be shown (e.g. see \cite{dm})
that the closed string twist sector modes couple
to Fayet-Iliopoulos and other linear terms in the D-brane world-volume
Lagrangians.  The details depend on $p$ but in general
the $(c,a)$ fields (geometrically, corresponding to a K\"ahler modulus)
$\omega_g$ in the sector twisted by the conjugacy
class of $g$ produce the Fayet-Iliopoulos terms
$\int d^4\theta\ \zeta_a V_a$ with
$\zeta_a = \sum_g \Tr r_a(g) \omega_g$.
Here $V_a$ is the vector superfield for $U(1)\subset U(n_a)$.
This provides independent FI terms for all $U(1)$'s with charged
matter, i.e. all but the overall diagonal $U(1)$.
Similarly complex structure moduli produce linear terms in the
superpotential.

These modifications will deform the D and F-flatness conditions
(schematically) as
\eqn
[Z,Z] = \zeta
\enq
and thus deform the moduli space -- strongly for $|Z|^2\sim \zeta$,
but negligibly for $|Z|^2 >> \zeta$.  The result for the regular
representation is a moduli space asymptotic to $\BC^n/\Gamma$ but
which -- for generic $\zeta$ -- turns out to be smooth and
topologically non-trivial.  The metric on this moduli space is an
example of a D$0$-brane metric $g^D$.
Physically one has realized this non-trivial moduli
space by embedding it in a larger but trivial configuration space,
as the minima of a linear sigma model potential.  The additional
degrees of freedom have masses of order $m \sim \zeta^{1/2}/l_s^2$
and thus one also sees an example of the bound $E << l_R/l_s^2$
we mentioned earlier.

In general one has an interesting quantum theory to solve at this
point, but in the case of D$0$-branes with blow up parameters which
keep us in the substringy regime $l_s^2 >> \zeta >> l_p^2$ the quantum
corrections are subleading and we can compute the
effective action by classical methods. In particular, computing
the metric on moduli space reproduces the symplectic quotient
construction of K\"ahler and hyperkahler metrics.\cite{hitchin}

A simple way to do this computation is to first write out the action in
$\CN=1$ 
superfield notation and substitute a solution of the F-flatness conditions.
One then fixs the complexified gauge invariance present before fixing
Wess-Zumino gauge in some explicit way.  Finally one minimizes the
action with respect to the vector superfields, considered as complex
variables.  This procedure combines the D-flatness conditions with the
fixing of ordinary gauge symmetry to provide an explicit K\"ahler
potential for the quotient metric. 

Let us do $\BC^2/\BZ_2$ (following \cite{dg}).
The regular representation is $n_1=n_2=1$; dropping the diagonal $U(1)$
we have $U(1)$ gauge theory with
four chiral superfields $b_0$, $b_1$, $\bar b_0$ and $\bar b_1$
with charges $+2$, $-2$, $-2$ and $+2$ respectively.
The gauged K\"ahler potential is then
\eqn
K = e^{2V} (|b_0|^2 + |\bar b_1|^2) + e^{-2V} (|b_1|^2 + |\bar b_0|^2 )
        - \zeta V
\enq
while the F flatness condition is
\eqn
b_0 \bar b_0 - b_1 \bar b_1 = 0.
\enq
(In general a complex structure modulus appears on the right, but we
set this to zero.)

We first relabel the variables $q\equiv e^{2V}$, $b_0=z$, $\bar b_0=w$.
We then make the gauge choice $\bar b_1=1$.  This
makes the F-flatness condition easy to solve, $b_1=zw$.
We then have
\eqn
K = q(1+|z|^2) + {1\over q}(1+|z|^2)|w|^2 - \zeta\log q.
\enq
The condition $\p K/\p q=0$ determines
\eqa\label{ehint}
0 &= q^2 - |w|^2 - {\zeta q\over 1+|z|^2}  \cr
q &= {1\over 2(1+|z|^2)}\left(\zeta \pm \sqrt{\zeta^2+4(1+|z|^2)^2|w|^2}\right)
\ena
and after substituting back,
\eqn
K = \pm \sqrt{\zeta^2 + 4(1+|z|^2)^2|w|^2} - \zeta \log q
\enq
with $q$ as in (\ref{ehint}).
The result is the K\"ahler potential for the Eguchi-Hanson metric,
a Ricci flat metric asymptotic to $\BC^2/\BZ^2$, but with the
singularity resolved to an $S^2$ of volume $\zeta$.
Actually it was clear without computation that this would come out,
as the $\CN=2$, $d=4$ supersymmetry guarantees that the quotient metric
will be hyperkahler; this implies Ricci flatness
and in four dimensions self-duality.  
This construction generalizes to
the general ADE case (though the D and E metrics require nonabelian
quotients and have not been worked out explicitly to my knowledge).

We can repeat the same computation for $\BC^3/\BZ_3$.  The regular
representation is $n_1=n_2=n_3=1$ and this is a $U(1)^2$ gauge theory
with two Fayet-Iliopoulos parameters $\zeta_1$ and $\zeta_2$.
The computation is done in \cite{dg} and we can thus quote an
explicit substringy K\"ahler potential producing a D$0$-metric $g^D$:
\eqa\label{z3metric}
K &= \left(2 + {1\over \zeta_1-\zeta_2}\right)v
 - \zeta_1\log(v + \zeta_1) \cr
 &- \zeta_2\log(v + \zeta_2)
 + (\zeta_1 + \zeta_2)\log (1 + |x|^2 + |y|^2) ,
\ena
with
\eqn
v(v+{\zeta_1})(v+{\zeta_2}) = (1 + |x|^2 + |y|^2)^3 |z|^2.
\enq
This time there was no a priori reason for the metric to be Ricci flat,
and indeed it is not for $\zeta\ne 0$.  The topology is that the
singularity has been resolved to a $\BC\BP^2$ with K\"ahler class
$\zeta$ (this is $z=0$) and sectional curvature $1/\zeta$.

Actually there was a hidden assumption in the previous analysis which
we now must justify.  The metric resulting from the quotient
construction depends on the K\"ahler potential of the original world-volume
gauge theory -- at the orbifold point this will be $K_0 = \sum_i |Z_i|^2$,
but we should check whether this receives corrections at non-zero
$\zeta$.  We might expect this to be true
as for K\"ahler class $\zeta >> l_s^2$ the metric
must reduce to a Ricci-flat metric, and there should be a sign of this
even at small $\zeta$.
An explicit world-sheet computation is rather
similar to a direct computation
of $g^D$ in conformal field theory and it is not hard to see that
corrections can appear at $O(\zeta/l_s^2)$.

This question is even more interesting as there do exist
modified $K_0'$ for which the quotient
construction will produce a Ricci-flat metric.\cite{ray}
On the other hand, these are nonanalytic in $\zeta$ and $z$ and
this is not what we expect to get from conformal field theory,
which (since the orbifold point is non-singular)
will produce an effective action analytic in the fields.
Thus (\ref{z3metric}) is the leading term in an expansion in positive
powers of the curvature length $(l_R/l_s)^2$.

\subsection{The D$0$-D$6$ system}

The metric around a D-brane will also be substringy.  Many examples
have been worked out and a particularly simple one is the D$0$-D$4$
system.  This theory has eight supercharges and the
strings stretched between the branes live in a hypermultiplet;
integrating them out produces the expected 
(supergravity) D$4$ metric at one loop.
One can show using supersymmetry that this result is 
exact.\cite{diacentin}  Examples with less supersymmetry have no a priori
reason to agree with supergravity and an example which does not
(and deserves more detailed study) is given in \cite{dps}.

Another interesting example
with no apparent supergravity interpretation is
the D$0$-D$6$ system.  This is an interesting example because one
can easily work out the form of the interaction at all length
scales (at least at order $g_s$) and see a non-trivial crossover.
Also interesting is that it preserves no supersymmetry yet can be
analyzed perturbatively -- this is because the leading interaction
is a repulsive potential.

As is well known the supergravity interpretation of the D$6$-brane
is as the KK monopole of $d=11$ supergravity reduced to the \IIa\ theory.
From this point of view the long range fields must fall off as $1/r$.
This includes the dilaton and this means that the D$0$ world volume
theory will include a potential (the mass term 
${1\over g_sl_s}\int dt\ e^{-D}$).
At the core the radius of the eleventh dimension goes to zero,
corresponding to $D\rightarrow -\infty$ and an infinite repulsive potential.

On the other hand, in gauge theory terms the interaction is mediated
by light open strings coming from the Ramond sector (the NS strings are
massive).  These form a fermion doublet under the $SO(3)$ rotational
symmetry of the transverse dimensions and their Hamiltonian is simply
\eqn\label{spinham}
H = E_0 + \bar\chi \sigma_i \chi X^i = E_0 + S_i X^i .
\enq
where $S_i = \sigma_i$ acting on the two-dimensional quantum
Hilbert space of the fermions and $E_0$ is a constant shift of the
energy coming from the massive open strings.

The ground state energy of this Hamiltonian is $E_0-|X|$ and this
description also leads to a repulsive potential -- but with a different
short distance behavior.  Clearly these are two limiting behaviors of
a continuous function which crosses over at $|X|\sim l_s$ -- this
function has been computed in \cite{lifschytz} and is monotonic.

Another amusing aspect of this system is that it provides a very simple
way to check the Dirac quantization of the RR charges of the two branes.
\cite{berkdoug}
The D$0$ sees the magnetic RR potential produced by the D$6$ as a Berry
phase associated with the Hamiltonian (\ref{spinham}), and it is easy
to show that this phase is given by the unit magnetic monopole potential.

\subsection{Problems involving several D$0$-branes}

As is well known the moduli space metric for configurations of several
static D$0$-branes is flat, due to cancellations between graviton,
RR vector and dilaton exchange.  This has a simpler eleven-dimensional
origin in the fact that pp-wave solutions exist with an arbitrary momentum
distribution in the transverse dimensions.

As is also well known by now the leading interaction 
between a pair of D$0$-branes in an expansion in
relative velocity is non-zero and takes the form $c v^4/r^7$.
This is also the precise result at $O(v^4)$ for the induced
potential computed in $N=16$ supersymmetric quantum mechanics,
the description valid at distances $r << l_s$, as can be understood
from supersymmetry -- massive string modes come in long multiplets
and do not contribute to this amplitude.  In fact supersymmetry
implies an exact non-renormalization theorem. \cite{sethi}

The Matrix theory proposal of BFSS leads to the conjecture that
an infinite series of generalizations of this apply, which predict
that the $O(v_1^2 v_2^2 \ldots v_k^2)$ effective potential for
$k$-graviton interactions will also reproduce classical supergravity results.
The original conjecture stated this for the limit in which gravitons are
described by large $N$ bound states of D$0$-branes, a limit which is as
yet not accessible to direct computation.  Although logically not
required, the simplest way this could
work is if the equivalence held at finite $N$ as well.
(See \cite{matrix} for discussion of this issue.)
For historical reasons this is generally known as the ``DLCQ conjecture.''

Some impressive computations have been done at two loops, in $SU(2)$
and $SU(3)$ gauge theory, which establish this conjecture for
three-graviton interactions.\cite{dlcq}
However this is not really convincing evidence for
the general conjecture, because these amplitudes are
strongly constrained by supersymmetry.  
In particular, using an argument given in \cite{martinec},
one sees that the known generalizations of supergravity 
with $16$ real supersymmetries -- the type I and heterotic
strings -- have exact three-graviton scattering amplitudes identical with
those of pure supergravity.  
Thus it appears quite possible that
supersymmetry and very weak consistency conditions force this result.

The four-graviton amplitudes come with $l_s$ corrections in all string
theories, so we know that modifications of supergravity exist at this
order.  Thus finding agreement here would constitute strong evidence
for the DLCQ conjecture, and conversely finding disagreement might give us
ideas for a new conjecture.  Straightforward three-loop computations
look formidable but not unimaginably difficult.  They are simplified
by introducing separations of scale (e.g. one D$0$ far from the others)
and in \cite{dine} some evidence for disagreement at three loops is given.

Another argument we could try to make for the DLCQ conjecture is the following.
We might imagine that whatever the results may be, they should be summarized
by an effective action coupling the stress-energy of the D$0$-branes
to a modified supergravity action; say
\eqa
S_{multi-D0} &= 
\int d^{10}x\ \sqrt{g} e^{-2D} \left( R + c l_s^6 R^4 + \ldots \right) \cr
&\qquad + \sum_{i=1}^N \int dt_i\ \delta(x-x_{0i}-v_i t_i)
 \left( e^{-D} g_{\mu\nu} v_i^\mu v_i^\nu + C\cdot v \right).
\ena
Although it is natural to suppose that this would be related to the action which
leads to the equations of motion (\ref{beta}) and thus $c\ne 0$,
I don't have a convincing argument for this, so let us regard it as
defined by computation.

In the substringy limit $r<<l_s$, one can see from the relation to gauge 
theory that the dependence on $l_s$ factors
out into the overall coupling constant.  Thus we appear to require a
modified supergravity action whose leading term is the Einstein action,
but with corrections independent of any additional dimensionful parameters.
If we knew that $S$ was generally covariant and local, no such corrections
would be possible.  

Conversely, it seems likely that if the DLCQ conjecture is false,
the substringy limit of $S_{multi-D0}$ will be non-local.
This is what we might expect intuitively as it is a low energy
effective action derived by integrating out strings of size $l_s$.
Despite its non-locality such a hypothetical new action might be
considered just as fundamental as the standard supergravity action,
and it would be extremely interesting to find an explicit description
of it.  

Another variation is to
combine our two problems and consider D$0$-brane interactions
in a curved background.  The cases which have
been studied so far are the leading $v^4$ interactions of two D$0$-branes
in a K3 or $C^2/\Gamma$ ALE background \cite{dos} and in a D$4$-brane
background.\cite{dd}  In both cases it is fairly easy to see that
the interactions differ drastically from those of supergravity and
the leading effects in inverse powers of the separation $r$ do not even 
behave as $1/r^7$ as would be predicted by a local action.
This certainly falsifies the DLCQ conjecture in curved space and
may even be evidence against the original conjecture; however it does
not disprove it at present as we do not know that a universal action
must exist in which both modifications of the original background
$g^{(0)}_{ij}$ and the D$0$-brane sources enter in the same way.

This result has been extended to D$2$-brane interactions in a very
pretty work \cite{dBHO}.  Following \cite{pp} one can compute
the leading interaction exchanging $p_{11}$ momentum as a $2+1$-dimensional
instanton process.  This instanton is the 't Hooft-Polyakov solution
(a $3+1$ monopole) and the scalar fields vary in a way which can see
the induced ALE geometry as the minimum of the potential.
Despite this qualitative agreement one can
show that the scalar field kinetic energy also contributes and
the interaction still does not behave as $1/r^7$.

After all this one might wonder whether there is some universal
obstruction to ever reproducing supergravity from finite $N$ gauge
theory in a curved background.  In some cases, it seems that 
there is.  The most basic condition for reproducing the $1/r^7$
interaction is that the mass of the degrees of freedom responsible
for it (the strings stretched between the interacting pair of
D-branes) be $m=r$.  One can study the problem of writing gauged sigma
model Lagrangians for which this is true and in
the specific case of a Calabi-Yau three-fold target which is not
Ricci flat (as we expect at substringy scales) no such action with
the required $\CN=1$, $d=4$ supersymmetry exists.\cite{dko}

\subsection{Conclusions}

To summarize our first lecture -- there are a wide variety of D-geometry
problems which provide a new and fascinating generalization of general
relativity and supergravity.  
In general these problems have both ``stringy'' and ``quantum'' aspects.
For example, one might consider
spacetime-filling D$3$-branes at points in the internal manifold; the
results we discussed would provide a classical limit of the space-time
effective action.
By considering D$0$-branes we were able to 
emphasize the new stringy aspects.

Particular examples can be studied by using
the by-now familiar approach of identifying new light 
degrees of freedom appearing
at singularities, which dominate the substringy physics.
However this leads to a case by case analysis and the common origin of
these problems in string theory suggests that a more unified framework
can be developed.  One approach is through non-linear sigma models
and it seems quite likely that there are universal equations of motion
which govern possible D-geometries, perhaps the beta function equations
in the minimal subtraction scheme.  Since they determine a physical
observable, it is not unreasonable to hope that the exact equations have
some simple description.

Ultimately, we might hope to base D-geometry on
some analog of the general principles from
which general relativity and supergravity were derived and understood.
Such principles probably will not be formulated in the context of
either limit $l_R >> l_s$ or $l_R << l_s$ but instead assume finite
$l_s$, as do algebraic approaches to conformal field theory.

\section{What is noncommutative geometry}

Much of mathematics is an interplay between geometry and algebra, 
two very different ways of thinking about the same objects.

Of the many algebras we can associate with a space, the most fundamental
is the commutative algebra of functions on the space.  Given a
commutative algebra $A$, one is often better off thinking about the 
corresponding space.  This is particularly true for algebras over
$\BC$.  Simple axioms for $A$ can often guarantee that the space is ``nice.''
For example, if $A$ is a commutative $C^*$ algebra (with a unit), 
it is isomorphic
to the algebra $C(X_A)$ of continuous complex functions on some
compact space $X_A$.  The space is just the space of homomorphisms $\chi$
from $A$ to $\BC$ such that $\chi(1)=1$.

Noncommutative geometry is a broad topic in which one tries to do the
same for general algebras.  The first examples to think about are the
$N$-dimensional matrix algebras $M_N(A)$ where $A$ is a commutative
$C^*$ algebra, i.e. matrices of functions on some $X_A$.
For $N>1$ these algebras have no non-trivial homomorphisms to $\BC$,
so the previous strategy fails; however we clearly want to associate
to these algebras both the space $X_A$ and the number $N$ as interesting
topological data.

We can do this by considering projections, elements $e\in A$ such
that $e^2=e$.  Clearly these will be matrix functions which at every
point can be diagonalized with eigenvalues $0$ and $1$.  The number of
unitary equivalence classes of these determines $N$.  Given an $e$ of
rank $1$, the subalgebra
of elements $eMe$ is isomorphic to $A$, from which we recover $X_A$.

Projections are also useful in formulating a generalization of
the idea of vector bundle.  A vector bundle $V$ over $X_A$ is determined by the
algebra of its sections $S_V$.
$S_V$ admits an action by multiplication by functions on $X$; in other
words it is a module over $A$.  We can turn this around by considering
all modules over $A$ satisfying certain conditions, which will enable
them to determine vector bundles.  The condition is that $S_V$ is
free and projective: it can be obtained by taking $n$-component
vectors with components in $A$, and applying a projection in $M_n(A)$.

One can go quite far in reformulating topological notions in this way.
For example, one can generalize the formalism of characteristic
classes to this case, and prove index theorems for operators acting on
vector bundles over noncommutative spaces defined in this sense.\cite{connes}

There has also been much work in formulating geometric notions.
What is quite intriguing and useful is that much of this work 
takes as the central
concept gauge theory on a noncommutative space, in other words a
theory of connections on projective modules which allows writing an
action, a Dirac operator and so forth.  Combining this with the idea that
in D-brane physics, coordinates are naturally promoted to matrices,
we are motivated to study the subject with this physical application
in mind.

\subsection{Deformation quantization}

It is probably more helpful at this point
to give examples than to develop abstract formalism, and a broad class of
examples come from trying to 
deform algebras of commutative functions into noncommutative 
algebras.\cite{flato}

In other words, given functions $f, g\in C(X)$, 
we seek a new multiplication law
$f * g$ which is associative and reduces to the original one in some limit:

\eqn\label{deformedmult}
f * g = fg + {\hbar\over 2} \{f,g\} + O(\hbar^2)
\enq

Here $\hbar$ is just a parameter characterizing the limit and $\{,\}$
some bilinear operation characterizing the deformation to first order.
Imposing associativity will constrain this operation and determine
the $O(\hbar^2)$ and higher terms.
The problem is to find all such deformations,
and was solved in \cite{kontsevich,fedosov}.

One type of deformation which is not so interesting is to make
linear redefinitions of our functions.  In other words we choose an
invertible linear operator $O$, say with an expansion in $\hbar$ such as
\eqn
f \rightarrow Of \equiv f + \hbar v^\mu(x){\p f\over\p x^\mu} 
+ O(\hbar^2 \p^2 f).
\enq
We then define a new multiplication
\eqn
f * g \equiv O^{-1}(O(f) O(g)) .
\enq
Clearly this is not going to give us anything essentially new, 
and we should look for deformations modulo this gauge-like ambiguity.

After fixing this ambiguity, it can be shown that associativity
requires $\{,\}$ to be a Poisson bracket, and then determines the
higher order terms uniquely.  (The explicit expressions require making a
choice of connection compatible with the Poisson bracket; this choice
is related to the ambiguity).

In this sense of a series expansion
-- formal because we are not guaranteed that the series converges,
and in many examples it does not -- the problem
has been solved.  Deformation quantization gives us a
noncommutative algebra for each choice of Poisson structure 
on our manifold.  This choice can be parameterized by a bivector
$\Theta^{\mu\nu}(x)$ satisfying the Jacobi identity, or when this
is non-degenerate by a closed two-form $\omega=\Theta^{-1}$.

\subsection{The noncommutative torus}

An example where the series converges and this works perfectly is
to start with $X=T^n$ the $n$-torus.  Let this be coordinatized
by $x^i$ with $1\le i\le n$ and $0\le x^i < 2\pi$.
We take constant
$\Theta^{ij}$ and define the multiplication law as
\eqn\label{nctorus}
(f * g)(x) = \exp \left( -i\pi\Theta^{ij}{\p\over\p x^i}
{\p\over\p y^j} \right)\ f(x) g(y) \bigg|_{y=x}
\enq
This is the familiar rule for multiplying Weyl ordered symbols
of quantum mechanical operators.  We can present the same rule
in terms of generators and relations: any function can be written
as a sum of monomials in the variables
\eqn
U_i \equiv e^{i x^i},
\enq
and on these (\ref{nctorus}) becomes
\eqn
U_i * U_j = e^{\pi i\Theta^{ij}} e^{i(x^i+x^j)} 
= e^{2\pi i\Theta^{ij}} U_j * U_i .
\enq
This gives us a purely algebraic definition and
proves that the deformed multiplication is associative.
It also shows that the noncommutative tori associated with
$\Theta^{ij}$ and $\Theta^{ij}+\Lambda^{ij}$ with $\Lambda^{ij}$
any antisymmetric matrix with integral coefficients are in fact
the same.

\subsection{Fuzzy spaces}

A different sense in which a general algebra can approximate the algebra of
functions on a continous manifold is to propose a {\it sequence} of
algebras $A_N$ and linear maps $\phi_N: C(X) \rightarrow A_N$
such that, as $N\rightarrow\infty$, we approximate the conventional
multiplication law arbitrarily well:
\eqn
\phi_N(f) \phi_N(g) = \phi_N(fg) + O({1\over N}) .
\enq
For example there is a well-known mapping of functions on $S^2$ into
$M_{2J+1}(\BC)$ which works by identifying the embedding
$S^2\subset \BR^3$ defined by $V = X_1^2 + X_2^2 + X_3^2 - 1 = 0$ with the
relation $L_1^2 + L_2^2 + L_3^2 = J(J+1)= (N^2-1)/4$ 
satisfied by the generators
of the $SU(2)$ representation of spin $J$.
More explicitly, a given $f$ on $S^2$ determines $f$ on $\BR^3$ up to
adding an arbitrary function multiplied by $V$; thus if we replace
$X_i \rightarrow L_i/\sqrt{(N^2-1)/4}$ we get a unique element of 
$M_{2J+1}(\BC)$.  The noncommutativity $[X_1,X_2] \sim X_3/N$ and goes
away for suitable test functions in the limit.

These constructions also have many physical applications (such as
the membrane construction of \cite{dhn}; see \cite{madore} for others)
but it is worth keeping in
mind the essential distinction between algebras which are ``as large''
as the space of functions on some starting manifold (like the
deformation quantization) and algebras which are smaller or even finite
(see \cite{rieffel} for a precise version of this distinction).
Indeed the two ideas seem somewhat orthogonal and one can pose
for example the problem of finding sequences of finite
dimensional algebras which approximate
(\ref{deformedmult}) for some finite $\hbar$.  

Having said this, we will return to algebras which are deformations
of function algebras.

\subsection{Gauge theory on the noncommutative torus}

Clearly to define gauge theory we want notions of derivative, integral
and metric on our noncommutative space.  Although general definitions
have been proposed, working them out for deformation
quantization of curved manifolds seems to be tricky
(e.g. see \cite{garcia}).
Of course the torus is simple: we can take
\eqa
\delta_i U_j &= i\delta_{ij} U_j \cr
\int U_1^{p_1}\ldots U_n^{p_n} &= \delta_{p_1,0} \ldots \delta_{p_n,0}
\ena
and use a constant matrix $g_{ij}$ as our metric.

The simplest definition of the gauge theory then proceeds
as follows.  First, take the
large $N$ limit of Yang-Mills theory -- the fields
$A_\mu$, $X^i$ and $\chi$ become large matrices.  At a generic
point in their configuration space, they satisfy no relations --
therefore
if we write a formula using the product $A_1 A_2$, then $A_1 A_2$ it
must stay throughout our formal manipulations.  Therefore gauge
invariance, supersymmetry, and all the other classical properties of
large $N$ Yang-Mills theory must be equally true if we regard the
fields as taking values in an arbitrary noncommutative algebra.

Such an approach will give an explicit gauge theory Lagrangian which
we can use to quantize, derive Feynman rules, and so forth.
In terms of the multiplication law (\ref{nctorus}), one essentially
finds the conventional Yang-Mills Lagrangian where every appearance
of the commutator $[A_\mu,A_\nu]$ is replaced by the algebra
commutator $A_\mu * A_\nu - A_\nu * A_\mu$.  In the case of pure
deformation quantization of functions (a ``$U(1)$'' theory) this
is essentially the Moyal bracket.  We thus find a non-local theory
of a very specific type, studied now in many works
(I found \cite{ncgaugeone,ncgaugetwo,ncgaugethree} especially illuminating).

One must also classify topological sectors of the gauge theory and
the formalism we just described is cumbersome for this purpose.
It is better to start from the beginning and classify projective modules
on the noncommutative torus, define connections on these
modules as derivations satisfying 
\eqn
\nabla_i (f v)=(\delta_i f)v + f \nabla_i v 
\enq
and then interpret these connections in terms of explicit gauge fields.
This is summarized for $T^p_\theta$ wit $p=2$
in \cite{cds} and for higher $p$
in \cite{rieffeltori} where a classification of constant
curvature connections is made.  

\subsection{Quantizing noncommutative gauge theory}

Although it is somewhat out of the main line of these lectures it
should be said that at present the most pressing question regarding this theory
is whether or not the quantum theory is well-defined.  The
non-locality of (\ref{nctorus}) looks very bad at first but after
deriving the Feynman rules and studying the one loop amplitude one
sees that the situation is actually rather good.

The relation to conventional Yang-Mills we described means that the
Feynman rules can be obtained by following the same procedure as for
conventional Yang-Mills with one of the standard gauge fixings, and
then replacing the Lie algebra structure constants $f_{abc}$ with
momentum-dependent structure constants coming from evaluating
(\ref{nctorus}) in momentum space.  At least in perturbation theory,
the gauge fixing procedure does not care about the nonlocality of
the interactions.
The result is
\eqn
f_{abc} \rightarrow 2i \sin \pi\Theta^{ij} p_{ai} p_{bj}
\enq
where $p_{a}$ and $p_{b}$ are the incoming momenta on these two
lines.  This expression is cyclically symmetric thanks
to momentum conservation $p_a + p_b + p_c = 0$.  Of course the
momenta are quantized on the torus.

Thus a one-loop two-point amplitude will take the general form
\eqn
\CA(p) = \sum_k {N(k,p)\over k^2 (k-p)^2 } 
\sin^2 \pi\Theta^{ij} k_{i} p_{j} .
\enq
where $N(k,p)$ is a standard polynomial in the momenta.
In the supersymmetric case,
one can use conventional superfield techniques as well.  For example,
the one-loop contribution to the kinetic terms will be given by this
expression with $N=1$.

What we see is that the extra factors associated with
the non-locality of the interaction do not worsen the UV behavior of
perturbation theory and actually can improve it, by providing
oscillatory factors.  In the example at hand one sees that there is
a term with the conventional Yang-Mills UV behavior (leading to a
non-zero beta function in $d=3+1$) and a new term in which the
oscillatory behavior cuts off the integral at the scale 
$\Lambda^2 \sim (\Theta p)^2$.  This leads to a finite term
$p^2 \log p^2$ in the one-loop effective action which in itself
is not inconsistent (since we are on the torus, $p$ is discrete and
we need not address the interpretation of the $p\rightarrow 0$
singularity).  It is quite plausible that the usual Ward identities
for gauge invariance will go through leading to one-loop
renormalizability.  This claim for the fermionic contribution is
already implicit in the literature -- it just says that the
log divergent part of $\det \Dslash$ is proportional to the NC
Yang-Mills action, a point made in \cite{connes}.

However the appearance of new non-local but finite contributions at one
loop suggests the possibility that similar divergent terms could be
produced at higher loop order.  Presumably a divergence proportional to
$p^2 \log p^2$ for example would spoil renormalizability (unless such terms
sum up to something sensible at finite $g$, in which case it would presumably
indicate that the UV structure of the theory is quite different from
conventional YM).
Since the theory is non-local, we do not have the standard
renormalization group arguments to guide us, and 
it is important to understand this point.

For the particular case of noncommutative $\CN=4$ SYM, we can be fairly
hopeful.  
First, since conventional $\CN=4$ SYM is finite, to the extent that amplitudes 
are given by conventional Feynman diagrams plus more convergent corrections,
it will also be finite.  Furthermore, the theories
appear as brane world-volume theories in limits of
string theory, as we discuss below,
so if they decouple from the rest of the string
theory at energies low compared to the string scale,
we have another strong argument for consistency.

\subsection{Quotients}

Actually we already made a noncommutative geometry construction in our
first lecture, in writing (\ref{orbifold}),
\eqn
\gamma^{-1}(g) \phi \gamma(g) = R(g) \phi \qquad\qquad 
{\rm for\ all}\ g \in \Gamma.
\enq
This is a universal
definition one could use to define the quotient of {\it any} algebra
(with some unitary representation)
by {\it any} group action.  As we saw it produces larger configuration
spaces than our starting theory and for this reason this construction
is called the ``crossed product'' in the math literature.

An attractive aspect of this construction, much discussed in \cite{connes},
is that it allows defining quotients which would be extremely singular if
we tried to use the simpler definition $f(x)=f(gx) \ \forall\ g\in\Gamma$.
The classic example is an ergodic action of a group, for which the orbit
of a point is dense in the whole set.

The simplest example is to take $x \in S^1$, say $0\le x< 1$, and
consider the action by $\BZ$ generated by $g(x) = x + \theta$ for
some constant $\theta$.  For $\theta=p/q$ rational of course the
usual quotient makes sense and produces a circle of radius $R/q$.
For $\theta$ irrational the only continuous functions on this quotient
are constant so we must consider the new definition.  

The regular representation of $\BZ$ is of course $g\ket{n}=\ket{n+1}$
and thus we can write solutions of (\ref{orbifold}) as matrices
$f_{mn}(x)$ satisfying
\eqn\label{torusid}
f(x)_{m+1,n+1} = f(x+\theta)_{m,n}.
\enq
For every function $f$ there is a solution $M_f$, the diagonal
matrix $f_{m,n}(x) = \delta_{m,n}f(x+n\theta -1)$.
A new solution of (\ref{torusid}) is
\eqn
S_{m,n} = \delta_{m+1,n}
\enq
as well as any power of $S$ or product $S^k M_f$.

This is somewhat like the familiar construction of D-branes on $S^1$ from those
at a point \cite{taylor} but here our starting point was already
a gauge theory on $S^1$, so what do we end up with?
In both cases we can interpret the expansion of the solution
in powers of $S$ as the Fourier expansion of a function in one 
higher dimension.
Now we already started out with one dimension, which we can also Fourier
expand writing $f(x)=\sum_n f_n T^n$ with $T=e^{2\pi i x}$.
Let us for later use slightly generalize to $S^1$ with circumference $L$,
for which $T=e^{2\pi i x/L}$.

The result is that the quotient algebra is generated by the two operators
$S$ and $T$ satisfying
\eqn\label{nccomm}
S T = e^{2\pi i \theta/L} T S.
\enq
The quotient in this sense of $S^1$
by a shift $\theta$ is the noncommutative
torus $T^2_{\theta}$.

Note that even for rational $\theta$ the two definitions of quotient
disagree.  We will see in the next section that
$T^2_{p/q}\cong M_q(T^2)$ so there is a close relation between them.
A useful description of this relation is that they are ``Morita equivalent.''
One way to think about this equivalence is that the two algebras must become
isomorphic upon tensoring with matrices $M_N(\BC)$ in a suitable large $N$
limit (more precisely, with the algebra of compact operators).

A stronger and more useful relation would be that there exist linear
maps between the two algebras which we could use to show that the gauge
theories built on the two algebras are isomorphic -- any configuration
in the gauge theory built on $A$ has a corresponding configuration in $B$ with
the same action.  Such a relation of ``complete Morita equivalence'' has
been defined and studied in \cite{schwarz}.  It would
imply not just that the BPS spectra in the two theories are the same 
but presumably that the theories are identical even classically, making
duality between them manifest.

\subsection{Duality on the noncommutative torus}

It is clear that the algebras $T^2_\theta$ and $T^2_{\theta+1}$ are
isomorphic.  We now argue that the algebras $T^2_\theta$ and
$T^2_{-1/\theta}$ are Morita equivalent.

This can be done by repeating the construction of the previous section
but by taking as the starting point an algebra defined using the
quotient $S^1=\BR/\BZ$.  Thus we start with functions on $\BR$ and again
find all solutions of the relation (\ref{torusid}).  These are again
the operator $S$ and $M_f$, but now with no periodicity imposed on $f$.

The algebra generated by these operators is clearly larger than that
of functions on $S^1$ with radius $\theta$.  
However according to general theory\cite{connes} the
two are Morita equivalent which is good enough for present purposes.

A heuristic argument for this is to imagine starting not with $\BR$
but with $S^1$ of radius $L=N\theta$ for some integer $N$.  By the previous
section the result is the NC torus with parameter $1/N$ -- if we
expand functions $f$ in terms of $T=\exp 2\pi i \sigma/L$, we have
(\ref{nccomm}) with $\theta/L=1/N$.
Now we rewrite the expansion in terms of $U \equiv T^N$ as
\eqn
f = \sum_{m=0}^{N-1} \sum_n T^m U^n f_{m,n} .
\enq
The operator $U$ generates the commuting algebra $A_{S^1}$ 
of functions on $S^1$
with radius $\theta$, while the pair $\{S,T\}$ as is well known have an
explicit $N$-dimensional ``clock and shift'' representation.  Thus the
result is equivalent to $M_N(A_{S^1})$, and it should be plausible that
the $L\rightarrow\infty$ limit produces the large $N$ limit of our
previous discussion.

Let the radius in the first quotient be $\theta_1$;
we can then take a second quotient by $x\rightarrow x+\theta_2$ by
repeating the discussion of the previous section, to produce the
algebra of matrices built on the noncommutative torus $T_{\theta_2/\theta_1}$.

Now the point of all this is that we made the two quotients in exactly
the same way -- so we can do them in the opposite order to get another
Morita equivalent algebra.  This will be the
algebra of matrices built on the noncommutative torus 
$T_{\theta_1/\theta_2}$.\footnote{
Actually we wanted $-\theta_1/\theta_2$.  At the level we are working we
cannot tell the difference as the
two are related by redefinition $T\rightarrow T^{-1}$.}

The upshot is that the symmetries $\theta\rightarrow -1/\theta$ 
and $\theta\rightarrow\theta+1$ have a 
``geometric'' origin very analogous to the familiar $\tau\rightarrow -1/\tau$
of elliptic curves.  We can define the NC torus as the quotient of $\BR$ by a
two-dimensional lattice, whose natural automorphism group is $SL(2,\BZ)$.

\subsection{Relation to M theory and prospects}

The applications of the noncommutative torus to M theory at present 
follow by combining existing results and dualities with
the following: if we consider D$1$-branes on a highly oblique torus, their
world-volume theory is $d=2+1$ noncommutative gauge theory.\cite{dh}
The main idea is expressed in figure 1.
\begin{figure}
\epsfxsize2.5in\epsfbox{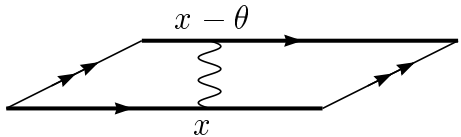}
\caption{A string excitation on a D1-brane (the thick line) on $T^2$.  
It is a bilocal object
and the resulting non-locality is precisely described by the NC gauge theory.}
\end{figure}

By duality this tells us that D$0$-branes on a very small torus with
non-zero Neveu-Schwarz two-form field are also described by noncommutative
gauge theory.  In the large $N$ limit this becomes the
Matrix theory description of M theory with a non-zero background
three-form $C_{-ij}$.\cite{cds}

The duality we found in the previous section then has the natural
interpretation as the zero volume limit of the usual string theory
$SL(2,\BZ)$ duality acting on $B+iV$.\footnote{
The connection with zero volume limits was recognized in the
early math work leading to noncommutative spaces being called ``tiny
spaces'' at first.  However it was soon realized that this name
could be a barrier to the subject being taken at all seriously and
the present name was adopted instead.\cite{connesremark}
}
The element $B\rightarrow -1/B$ thus is double T-duality
(in Matrix theory it is double T-duality including the compactified
null direction), which
is not a geometrical symmetry in the usual sense of supergravity.
We see that it has become a geometrical symmetry in this 
framework.  
A test of this idea is that noncommutative $T^d$ should have $SO(d,d;\BZ)$
symmetry and indeed it does, as
has been shown by Rieffel and Schwarz.\cite{riefschw}

In four dimensions, a very interesting question is to understand the
structure of the moduli spaces of self-dual NC gauge fields.
The first work on this was \cite{nekschw} which studied instantons
on $\BR^4$ and showed that they can be obtained from a
pretty noncommutative deformation of the ADHM construction: 
one simply takes the matrices $X^\mu$ to be noncommuting in the same
way as the original coordinates.  This smooths out singularities in the
moduli space and, in contrast to the standard case, even the $U(1)$
theory has smooth self-dual solutions.  

We can expect a similar story on tori -- moduli spaces will be deformations of 
those of conventional gauge theory -- although details have not yet been
worked out yet.  It seems very plausible that these
moduli spaces will appear as moduli spaces of string compactifications,
and in fact the recent work \cite{ganor} uses string dualities to
argue that moduli spaces of $(2,0)$ theories in six dimensions compactified
on $T^3$ with Wilson lines coupling to the R-symmetry currents are moduli
spaces of instantons on NC tori.

It seems quite likely that NC gauge theory and much of the setup also
works on curved base spaces.  
Certainly the simple picture of figure 1 generalizes and
shows us that D-branes on general curved spaces will have a non-local
aspect to their world-volume theory which should be described by
noncommutative geometry.  Indeed a construction closely related to
the crossed product is that
of the foliation algebra and this should be useful in describing families of 
brane hypersurfaces.  This may lead to gauge theories produced
by deformation quantization, or it might lead to a better behaved version
of deformation quantization.

Asymmetric orbifolds, their deformations and other theories now lacking any
geometric interpretation are also natural candidates to try to cast in
this language.  Discrete torsion provides an example in which noncommutativity
plays a clear role,\cite{disc} and about which there is clearly more to say.

Finally, on a more speculative side,
it seems quite possible that the key to understanding
D-geometry will turn out to lie in a deeper understanding of 
noncommutative geometry.  D-geometry provides a set of problems in which
the non-locality of string theory and M theory -- still difficult for us
to treat -- is reduced to specific deformations of the traditional notions of
geometry, which can be described using formal series expansions in
the operator $l_s \partial$.  Deformation quantization involves an
analogous expansion and it is probably significant that in
Kontsevich's construction its coefficients are obtained
as correlation functions in an auxiliary ``string theory.''
Furthermore, compatibility 
with T-duality is known to lead to strong constraints on the sigma model RG 
\cite{haag}, so if noncommutative geometry can provide
a simple general description of T-duality, it seems quite likely to be an
appropriate language for this problem.

\end{document}